\documentclass[prl,aps,twocolumn]{revtex4}

\usepackage{epsf}

\begin{document}

\title{%
  \hfill{\normalsize\vbox{%
    \hbox{\rm March 2003} 
    \hbox{\rm DPNU-03-05}
    \hbox{\rm TU-685    }
  }}\\
  \vspace{-0.5cm}
  {\bf $\pi^+$-$\pi^0$ Mass Difference 
  in the Hidden Local Symmetry: \\
  A Dynamical Origin of Little Higgs}
}
\author{{\bf Masayasu Harada}$^{\rm(a)}$}
\author{{\bf Masaharu Tanabashi}$^{\rm(b)}$}
\author{{\bf Koichi Yamawaki}$^{\rm(a)}$}

\affiliation{$^{\rm(a)}$ Department of Physics, Nagoya University,
Nagoya, 464-8602, Japan,}
\affiliation{$^{\rm(b)}$ Department of Physics, Tohoku University,
Sendai, 980-8578, Japan.}

\begin{abstract}
We calculate $\pi^+$ - $\pi^0$ mass difference $\Delta m_\pi^2
\equiv m_{\pi^+}^2 -m_{\pi^0}^2$ in the Hidden Local 
Symmetry (HLS) model,
based on the Wilsonian matching and 
Wilsonian renormalization-group equations. Even without $a_1$ meson
the result agrees well with the experiment 
in sharp contrast to the
conventional approach where 
the $a_1$ meson plays a crucial role. 
For large $N_f$ QCD, there arises a large
hierarchy between $\Delta m_\pi^2$ and 
the $\pi$ decay constant $F_\pi^2$, $\Delta m_\pi^2/F_\pi^2 \ll 1$,
near the critical point where the chiral symmetry gets restored 
as the vector manifestation and the HLS model becomes a little Higgs
model 
with two sites and two links, with the dynamically generated gauge
coupling of the composite $\rho$ becoming vanishingly small.
\end{abstract}

\maketitle

The $\pi^+$ - $\pi^0$ mass difference $\Delta m_\pi^2
\equiv  m_{\pi^+}^2 -m_{\pi^0}^2$ was first successfully 
calculated~\cite{Das} by the
current algebra 
in conjunction with the Weinberg's 
spectral function sum rules~\cite{Weinberg}. 
Since then it has been a prototype of the
mass calculation of pseudo Nambu-Goldstone (NG) bosons 
in 
strong coupling gauge theories 
such as those 
in the technicolor theories~\cite{techni}
and more recently in the little Higgs models~\cite{ACG}.
Hence this type of calculation plays a central role of the model
buildings.

The basic technology to calculate those pseudo NG bosons up to the
present  
has been an ancient one through the Weinberg's first and second 
sum rules~\cite{Weinberg}
saturated by the  $\pi$, $\rho$ and $a_1$ meson poles.
Then the calculation heavily depends on the somewhat 
elusive broad resonance of $a_1$ meson whose 
mass however substantially deviates from the prediction of the
Weinberg's  
sum rules. The reason why the method remains so awkward is due to our
ignorance  
of the strong coupling dynamics of QCD and QCD-like theories and
their effective field theory. Then the calculation is also
challenging for 
theorists to construct the effective field theory of hadrons.

Recently two of the authors (M.H. and K.Y.) developed an effective
field 
theory at loop order based on the Hidden Local Symmetry (HLS)
model~\cite{BKUYY,BKY}:
The bare parameters of the HLS model
was determined by those of the underlying  QCD through the matching of
current correlators  of both theories at a certain scale 
$\Lambda \,(\simeq 1.1. {\rm GeV})$ which is the cutoff for the HLS
model  (``Wilsonian matching'')~\cite{HY:WM}.
Once the bare parameters of the HLS model defined at $\Lambda$
were so determined, we did uniquely 
predict the low energy hadron physics by the one-loop renormalization
group  
equations (RGEs) due to the $\pi$ and $\rho$ loops including quadratic
divergences 
(``Wilsonian RGEs'')~\cite{HY:duality,HY:WM}. The
results were  
in remarkable agreement with experiments. (For a detailed review of
the whole approach see Ref.~\cite{HY:PRep}.)
 
In this paper we shall apply the same method of HLS model to the
calculation of  
the $\pi^+$-$\pi^0$ mass difference $\Delta m_\pi^2$.
The method is straightforward and has essentially no
ambiguity once we fixed the $\Lambda$ which was already fixed 
to be $\simeq 1.1 {\rm GeV}$ in the 
previous analyses. 
Remarkably, we can successfully reproduce the experimental value 
{\it without introducing the $a_1$ meson}  whose mass 
is higher than  our matching scale  $\Lambda\simeq 1.1 {\rm GeV}$.

Moreover, there occurs {\it cancellation of 
the quadratic divergences} in 
$\Delta m_\pi^2$ 
arising from the $\pi$ and $\rho$ loops
which in the usual approach is to be canceled by the conspiracy
between 
the $\pi$, $\rho$ and $a_1$ mesons as required by the pole-saturated form
of the 
Weinberg's first sum rule. It was shown in Refs.~\cite{HY:WM,HY:PRep}
that
the {\it bare} Lagrangian of our HLS model,
when the photon and $\rho$ gauge couplings are switched off, is very 
close to the Georgi's vector limit~\cite{Georgi},
which corresponds to locality of the theory space of the little
Higgs model 
of two sites and two links, and hence the one-loop absence of
quadratic 
divergence takes place for the same reason as in the little Higgs.
So this type of {\it little Higgs is already realized in the 
real-life QCD
}\,!

Although
the dynamically 
generated HLS gauge coupling of the composite $\rho$ is rather strong,
$g^2(\Lambda) \gg 1$, in the
real-life QCD with $N_f =3$,
it was found~\cite{HY:VM,HY:PRep} that  
when $N_f$ is increased in the underlying QCD
so that the chiral symmetry is expected to 
get restored at certain critical value 
$N_f^{\rm crit}$~\cite{IKKSY:98,Appelquist-Terning-Wijewardhana},
the corresponding HLS model goes over to the Vector Manifestation 
(VM)~\cite{HY:VM}
where the $\rho$ coupling as well as the $\rho$ mass and $F_\pi$
becomes  
vanishingly small;
$g^2 \rightarrow 0$, $m_\rho^2/\Lambda^2 \rightarrow 0$ 
and  $F^2_\pi/\Lambda^2 \rightarrow 0$.
Then the VM will in fact
provides a toy model for the
dynamical generation of the little Higgs
models out of strongly interacting underlying gauge theories.
We shall also demonstrate a large hierarchy $\Delta m_\pi^2 /F_\pi^2
\ll 1$ 
near the VM point 
as desired in the little Higgs model
building. 
However, we do not attempt here to construct a realistic model for the
little 
Higgs. The quartic coupling as well as the Yukawa coupling is not
considered either.
We do instead demonstrate a concrete example for 
a possibility to dynamically generate a class of little Higgs
models, with the locality of the theory space explicitly broken only
by  {\it weakly coupled} gauge interactions, out of 
{\it strongly coupled} underlying gauge theories. 

Let us start with briefly explaining 
the HLS model and its loop calculations (For a detailed review
see \cite{HY:PRep}).
The HLS model~\cite{BKUYY,BKY} 
is an extension of the nonlinear sigma model based on the
$G_{\rm global} \times H_{\rm local}$ symmetry, where
$G = \mbox{SU($N_f$)}_{\rm L} \times 
\mbox{SU($N_f$)}_{\rm R}$  is the 
global chiral symmetry and 
$H = \mbox{SU($N_f$)}_{\rm V}$  the HLS whose gauge bosons are
identified with the $\rho$ meson and its flavor partners (to be denoted
as $\rho$ hereafter).
Here $N_f$ denotes the number of massless quark flavors in the
underlying 
QCD (We take $N_f = 3$ for the real-life QCD. See 
\cite{HY:WM,HY:PRep}.).
The basic dynamical variables in the HLS model are gauge bosons
$\rho_\mu=\rho_\mu^a T_a$  
of the HLS and two 
SU($N_f$)-matrix-valued variables $\xi_{\rm L}$ and 
$\xi_{\rm R}$
parameterized as
$
\xi_{\rm L,R} = e^{i\sigma/F_\sigma} e^{\mp i\pi/F_\pi}
$
which transform as 
$\xi_{\rm L,R}(x) \rightarrow \xi_{\rm L,R}^{\prime}(x) =
h(x) \xi_{\rm L,R}(x) g^{\dag}_{\rm L,R}
$,
where $h(x) \in H_{\rm local}$ and 
$g_{\rm L,R} \in G_{\rm global}$.
Here $\pi = \pi^a T_a$
denotes the NG bosons ($\pi$ meson and its flavor 
partners) associated with 
the spontaneous breaking of $G$ and 
$\sigma = \sigma^a T_a$ (with $J^{PC} =0^{+-}$)
the NG bosons absorbed into the (longitudinal) HLS gauge bosons $\rho$
(not to be confused with the scalar boson ``sigma'' in the linear
sigma model which has $J^{PC}=0^{++}$). 
$F_\pi$ and $F_\sigma$ are the relevant decay constants, with a ratio
$a$ 
defined by
\begin{equation}
a \equiv F_\sigma^2/F_\pi^2
\ .
\end{equation}
The covariant derivatives of $\xi_{\rm L,R}$ are defined by
$
D_\mu \xi_{\rm L} 
=
\partial_\mu \xi_{\rm L} - i g \rho_\mu \xi_{\rm L}
+ i \xi_{\rm L} {\cal L}_\mu 
={\cal D}_\mu \xi_{\rm L} - i g \rho_\mu \xi_{\rm L}
$,
and similarly for ${\rm L} \rightarrow {\rm R}$,
where $g$ is the HLS gauge coupling.
${\cal L}_\mu$ and ${\cal R}_\mu$ denote the external gauge fields
(such as the photon and $W$ and $Z$ bosons)
gauging the $G_{\rm global}$ symmetry.

The (bare) HLS Lagrangian at 
${\cal O}(p^2)$ is given by~\cite{BKUYY,BKY}
\begin{equation}
{\cal L}_{(2)} = F_\pi^2 \, \mbox{tr} 
\left[ \hat{\alpha}_{\perp\mu} \hat{\alpha}_{\perp}^\mu \right]
+ F_\sigma^2 \, \mbox{tr}
\left[ 
  \hat{\alpha}_{\parallel\mu} \hat{\alpha}_{\parallel}^\mu
\right]
+ {\cal L}_{\rm kin}(\rho_\mu) \ ,
\label{Lag 2}
\end{equation}
where ${\cal L}_{\rm kin}(\rho_\mu)$ denotes the kinetic term of
$\rho_\mu$. 
In the unitary gauge $\sigma=0$, 
the second term, containing
$
\hat{\alpha}_{\parallel}^\mu =
(  D_\mu \xi_{\rm L} \cdot \xi_{\rm L}^\dag +
  D_\mu \xi_{\rm R} \cdot \xi_{\rm R}^\dag 
)
/ (2i)
= {\cal D}_\mu \xi_{\rm L} \cdot \xi_{\rm L}^\dag +
  {\cal D}_\mu \xi_{\rm R} \cdot \xi_{\rm R}^\dag
  - g \rho_\mu 
$,
yields the $\rho$ mass term
$M_\rho^2=(g F_\sigma)^2$
as well as the $\rho\pi\pi$ coupling
$g_{\rho\pi\pi} =(a/2)g$,
$\rho-\gamma$ mixing $g_\rho=g F_\sigma^2$,  and the direct $4 \pi$
coupling, etc., while 
the first term containing
$
\hat{\alpha}_{\perp}^\mu =
 ( D_\mu \xi_{\rm L} \cdot \xi_{\rm L}^\dag -
  D_\mu \xi_{\rm R} \cdot \xi_{\rm R}^\dag 
)
=({\cal D}_\mu 
  \xi_{\rm L} \cdot \xi_{\rm L}^\dag -
 {\cal D}_\mu \xi_{\rm R} \cdot \xi_{\rm R}^\dag 
)/(2i)
$
is identical to the  usual nonlinear chiral Lagrangian based on $G/H$,
with 
$G$ being gauged by the external gauge bosons
${\cal L}_\mu$ and ${\cal R}_\mu$, 
where
the flavor 
chiral symmetry $G$ is given by the diagonal sum of $G_{\rm global}$
and $H_{\rm local}$, with the flavor vector symmetry $H$ being
 the diagonal sum of
$H_{\rm global} (\subset G_{\rm global})$ and $H_{\rm local}$.
In the low energy, $p^2\ll M_\rho^2$, where the $\rho$ kinetic term 
can be ignored,
the equation of motion of $\rho$ from the second term 
simply gives zero for the second term, thus the HLS model is reduced
to the first term, namely the usual (gauged) nonlinear chiral
Lagrangian 
based on $G/H$.

Let us now  
calculate $\pi^+$-$\pi^0$ mass difference or 
its $N_f$ generalization,
$\Delta m_\pi^2$, the mass of the pseudo-NG boson
associated with the $T_1$ generator
in the QCD with $N_f$ massless quarks.
The photon field $A_\mu$
reads $
{\cal L}_\mu = {\cal R}_\mu = e \, Q \, A_\mu 
$,
where $e$ is the electromagnetic coupling 
and $Q$ the electromagnetic charge matrix
of the diagonal form:
${\rm diag} (Q)=(2/3,-1/3, \cdot \cdot)$.
In order to include the photon loop, we need to add the 
kinetic term of the photon field to the 
${\cal O}(p^2)$ Lagrangian in Eq.~(\ref{Lag 2}).
The HLS Lagrangian further needs a 
{\it bare} term 
proportional to:
$
\alpha_{\rm em}
\, \Omega \,
\mbox{tr} \left[ Q U Q U^\dag \right]
$,
where 
$U=\xi_L^\dagger \xi_R=e^{2 i \pi/F_\pi(\Lambda)}$
and 
$\alpha_{\rm em} = e^2/4\pi$
is the fine structure constant.
The bare $\Delta m_\pi^2$ defined at $\Lambda$ is then given by
\begin{equation}
\left.
 \Delta m_\pi^2 
 \right
 \vert_{\rm bare}
= 
\alpha_{\rm em} \, 
 \Omega (\Lambda) 
 / F_\pi^2(\Lambda)\equiv
 \alpha_{\rm em} \,  \omega(\Lambda)
\ .
\label{bareomega}
\end{equation}

Such a bare term 
arises from integrating out the quark and gluon fields down to
the matching scale $\Lambda$ in the presence of
dynamical photon field and can be determined by the Wilsonian matching 
proposed in Refs.~\cite{HY:WM,HY:PRep}.
To estimate it, 
we rewrite~\cite{Yamawaki:82} the usual current 
algebra formula~\cite{Das}
for 
$\Delta m_\pi^2$
in terms of the full current correlators instead of the
spectral functions:
$\Delta m_\pi^2=
(3\alpha_{\rm em}/4\pi)
   \int_{0}^\infty d Q^2
\, Q^2 
\Delta \Pi(Q^2)/F_\pi^2(0)
$,
where $\Delta\Pi(Q^2)
\equiv \Pi_A(Q^2) - \Pi_V(Q^2)$ 
is the difference between the  axialvector correlator 
$\Pi_A(Q^2)$ and the
vector current correlator $\Pi_V(Q^2)$,
and
$F_\pi(0)(\ne F_\pi(\Lambda))$ the physical decay constant of $\pi$. 
Now we identify the high energy part of the
integral for $Q^2> \Lambda^2$
as the bare term Eq. (\ref{bareomega}):
\begin{equation}
\omega(\Lambda)=\frac{3}{4\pi}
\int_{\Lambda^2}^\infty d Q^2 Q^2 \frac{
\Delta \Pi^{\rm (QCD)}(Q^2)}{F_\pi^2(0)}
 = \frac{8 }{3 }
\frac{ \alpha_s \left\langle \bar{q} q \right\rangle^2 }
  {  F_\pi^2(0)\Lambda^2 }
\ ,
\label{match om}
\end{equation}
where
$\Delta \Pi^{\rm (QCD)}(Q^2)$ is given by 
the operator product expansion (OPE) in 
QCD~\cite{SVZ}:
$
\Delta \Pi^{\rm (QCD)}(Q^2) 
=
[4\pi(N_c^2-1)/N_c^2]
    [(\alpha_s \left\langle \bar{q} q \right\rangle^2)/Q^6]
$ and we set $N_c=3$.
Note that Eq.~(\ref{match om}) 
is positive and hence the {\it OPE gives a clear picture
that the QCD vacuum is aligned by the photon coupling in the desired
direction} as far as the bare $\omega$ is concerned.   

In the real-life QCD with $N_f=3$, 
Eq.~(\ref{bareomega}) with Eq.~(\ref{match om}) is estimated
as:
\begin{equation}
\left. \Delta m_\pi^2 \right\vert_{\rm bare}
 = \alpha_{\rm em}\omega(\Lambda)
=
211 \pm 47 \pm 140\, \mbox{MeV}^2 
\end{equation}
for a typical value of 
$\left( \Lambda\,,\,\Lambda_{\rm QCD}\right) = 
  (1.1\,,\,0.4)\,\mbox{GeV}$,
where the first error comes from $F_\pi(0)=86.4 \pm 9.7 \,{\rm MeV}$
(the value at chiral limit of $N_f=3$)~\cite{HY:PRep}
and the second one from 
$
\left\langle \bar{q} q \right\rangle_{\rm 1 GeV} =
- \left( 225 \pm 25 \, \mbox{MeV} \right)^3 
$~\cite{GL:PRep}.

Now we calculate 
one-loop contribution $\Sigma_{ab}$
(divergent part) to the $\pi_a$-$\pi_b$ two point function
from the photon loop in the HLS 
(For the Feynman rule see Ref.~\cite{HY:PRep}). 
In Landau gauge for the photon, 
the only relevant diagrams are a quadratically divergent 
$\gamma$ loop with the $\pi\pi\gamma\gamma$
vertex proportional to $(1-a)$, and 
a logarithmically divergent $\rho-\gamma$ loop 
(via $\rho-\gamma$ mixing) with 
the $\pi\pi\rho\gamma$ vertex, which is  proportional to
$ [a g^2 +(a-1)e^2]F_\pi^2\simeq a g^2 F_\pi^2=M_\rho^2$ (for
  $a g^2 \gg (a-1) e^2$):
$\Sigma_{ab} \vert_{\rm div}=
2 \, \mbox{tr} 
  \left[ [T_a \,,\, Q ] \, [ T_b \,,\, Q]
   \right]
 \alpha_{\rm em} 
 \left. \omega \right\vert_{\rm div}
 $,
where 
$
\left. \omega \right\vert_{\rm div}=
\frac{1}{4\pi}\left[
  (1-a) \Lambda^2 + 3 a M_\rho^2 \ln \Lambda^2
\right]
$~\cite{footnote1}.
Here we used as in Refs.~\cite{HY:duality,HY:WM,HY:PRep}
the dimensional regularization and identify the quadratic
divergences with 
the $n=2$ pole (Note that the coefficient of the quadratic divergence
is $1/3$ of that of the naive cutoff)~\cite{Veltman}.
The 
RGE for $\omega$
thus reads
\begin{equation}
\mu \, \frac{ d \omega }{ d \mu } =
- \frac{1}{ 2\pi } \left[
  ( 1 - a )\mu^2 + 3 a \, M_\rho^2 
\right]
\ .
\label{RGE om}
\end{equation}

We first solve  Eq.~(\ref{RGE om}), with the boundary 
condition Eq.~(\ref{match om}), from $\Lambda$ 
to $m_\rho$, with the physical mass
$m_\rho$ defined by
$
m_\rho^2 = M_\rho^2(\mu=m_\rho)=a(\mu = m_\rho) \, g^2(\mu=m_\rho) \,
F_\pi^2(\mu = m_\rho) 
$, 
which yields $\omega(m_\rho)$.
Here the RGEs of other parameters 
$F_\pi$, $a$ and $g$ 
were already solved 
in the previous analyses~\cite{HY:WM,HY:PRep}
in excellent agreement with the experiments,
with their bare values determined by the Wilsonian matching
of the HLS model with the underlying QCD through the OPE for the
current correlators.

At $\mu=m_\rho$
the $\rho$ gets decoupled, so that  
the RGE for $0<\mu <m_\rho$ should be changed to that 
of ChPT without $\rho$ loop where we 
change the notation of $\omega$ to $\omega^{(\pi)}$.
Then the RGE for  $\omega^{(\pi)}$
takes the form of that obtained by setting $a=0$ 
in Eq.~(\ref{RGE om}), 
which is readily solved
as
$
\omega^{(\pi)}(\mu) 
= 
\omega(0)
- \mu^2/4\pi
$
where 
$\omega(0) \equiv \omega^{(\pi)}(0)$.
Then we get $\omega(0)
=\omega^{(\pi)}(m_\rho) +m_\rho^2/4\pi$.
Actually, we needed to include finite renormalization effects to
match the HLS with ChPT in
the previous work~\cite{HY:WM,HY:PRep}.
Similarly to 
$F_\pi^2$
at 
$\mu=m_\rho$,
there exists a finite renormalization effect also for $\omega$:
Comparing the quadratic divergence of each RGE,
we have
$
\omega^{(\pi)}(m_\rho)=\omega(m_\rho) - 
a(m_\rho)m_\rho^2/4\pi
$.
Then,
\begin{equation}
\omega(0) = \omega(m_\rho) + 
 \left[ 1- a(m_\rho) \right]m_\rho^2/4\pi 
\ ,
\label{finite ren om}
\end{equation}
which  yields 
$\Delta m_\pi^2 =\alpha_{\rm em} \omega(0)$.

As shown in the previous works~\cite{HY:WM,HY:PRep},
the real-life QCD is close to the choice $a(\Lambda)\simeq 1$.
We thus first demonstrate a simplified analysis for an ideal case
$
a(\Lambda) = 1
$,
which was explicitly shown~\cite{HY:PRep}
to yield a reasonable agreement with
the $\rho$ and $\pi$ experiments: 
$F_\pi(0) = 73.6 \pm 5.7\,\mbox{MeV}$ (compared with
$86.4 \pm 9.7\,\mbox{MeV}$~\cite{HY:PRep}) 
and other quantities such as 
 $g_\rho$, $g_{\rho\pi\pi}$, $L_9$, $L_{10}$. 
Moreover,
in spite of the bare value $a(\Lambda) =1$,
the physical  
value  defined as $a(0)\equiv F_\sigma^2(m_\rho)/F_\pi^2(0)$
was predicted to be $\simeq 2.0$, very close to the successful
value of the tree-level phenomenology~\cite{BKUYY,BKY}. 
Note that the quadratic divergence for $\omega$ is proportional
to $(1-a)$ which is canceled for $a=1$ 
 without invoking the Weinberg's first sum rule.

Since $a=1$ is the fixed point
of the RGE~\cite{HY:duality,HY:PRep}, 
we have  $a(m_\rho) =1$
and hence $\omega(0)=\omega(m_\rho)$.
If we neglected the running of $M_\rho^2$ in Eq.~(\ref{RGE om}),
the RGE~(\ref{RGE om}) would be readily solved to give 
$\Delta m_\pi^2=\alpha_{\rm em} \omega(m_\rho)
=(3 \alpha_{\rm em}/4\pi)  M_\rho^2 \cdot
\ln (\Lambda^2/M_\rho^2) + \Delta m_\pi^2|_{\rm bare}$,
with $\Delta m_\pi^2|_{\rm bare}= 290\pm 149 \,{\rm MeV}^2$
(for $F_\pi(0) = 73.6\pm 5.7\, {\rm MeV}$ above), which
 would yield 
$\Delta m_\pi^2\simeq 1006 \,{\rm MeV}^2$
if we took $M_\rho^2$ as $m_\rho^2$~\cite{footnote2}.
Amazingly, even such a crude estimate is
in rough agreement with the experiment 
$\Delta m_\pi^2|_{\rm exp.} = 1261 \, {\rm MeV}^2$.
Actually, the running effect of $M_\rho^2 (\mu)$
boosts up 
the above quantum corrections: Solving Eq.(\ref{RGE om}) together 
with RGEs for other parameters as in \cite{HY:PRep}, 
we have 
\begin{equation}
\Delta m_\pi^2 = \alpha_{\rm em} \omega(0) 
=1223 \pm 263 \, {\rm MeV}^2\ ,
\end{equation}
for a typical case
$\left( \Lambda \,,\, \Lambda_{\rm QCD} \right) = 
(1.1\,,\,0.4)\,\mbox{GeV}$~\cite{footnote3},
where the error comes from the
$\langle\bar{q}q\rangle_{\rm 1 GeV}$
input.

Now in the full analysis of $N_f=3$ case~\cite{HY:WM,HY:PRep},
we used as an input the experimental value $F_\pi(0)=
86.4 \pm 9.7 \,{\rm MeV}$ instead of the ansatz  $a(\Lambda)=1$,
and predicted the low energy quantities in remarkable agreement 
with the experiments.
The bare parameter $a(\Lambda)$ in this case
was determined as
$a(\Lambda) \simeq 1.3 $
for $\left( \Lambda \,,\, \Lambda_{\rm QCD} \right) = 
(1.1\,,\,0.4)\,\mbox{GeV}$.
Under this full analysis setting, we compute $\Delta m_\pi^2$ as
\begin{eqnarray}
  \Delta m_\pi^2
  = 1129 \pm 18 \pm 218 \, \mbox{MeV}^2 \ ,
\end{eqnarray}
where the first error 
comes from the
$F_\pi(0)$
input. This is in good agreement with the experiment.

Thus we have successfully reproduced the experimental value of
$\Delta m_\pi^2$ in the HLS model with $a(\Lambda) \simeq 1$,
{\it without introducing the 
$a_1$ meson and without
invoking the  Weinberg's spectral function sum
rules}.

Now we discuss our result in connection with the little Higgs models.
We have seen that the real-life QCD is very close to $a(\Lambda) =1$,
which implies that the quadratic divergence of 
$\Delta m_\pi^2$ in the HLS model, Eq.~(\ref{RGE om}), does dissappear 
in accord with the little Higgs~\cite{ACG}. 
The HLS model with $a=1$ actually
 corresponds to the locality of
the theory space in the little Higgs models: When
the gauge couplings of both $\rho$ and photon are switched off,
$g=e=0$, 
the HLS Lagrangian takes the Georgi's vector limit~\cite{Georgi}
$G_1 \times G_2/G_{1+2}$ with 
$G= {\rm SU}(N_f)_{L} \times {\rm SU}(N_f)_{R}$,
which is nothing but a  little Higgs model with two sites and two
links.
This implies that the locality of the theory space is violated
only by the gauge couplings $g$  and $e$ even for the real-life QCD
with 
$N_f=3$: 
 $G_1$ is explicitly broken 
by the $\rho$ coupling down to $H_{\rm local}$ and $G_2$ 
becomes $G_{\rm global}$ of the HLS model, while $G_2$ (and hence
$G_{\rm global}$) is also explicitly broken by the photon coupling
down to ${\rm U}(1)_Q$, with those gauge symmetries spontaneously
broken in the Higgs mechanism: $H_{\rm local} \times {\rm U}(1)_Q
\rightarrow {\rm U}(1)_{\rm em}$. Then, as we have seen,
some of the NG bosons acquire a mass  
\begin{equation}
\Delta m^2_\pi \sim (3/4\pi) \alpha_{\rm em} m_\rho^2 
\sim (1/4\pi) \alpha_{\rm em} \alpha_{\rm HLS} \Lambda^2 
\label{PNGmass}
\end{equation}
(up to $\Delta m_\pi^2|_{\rm bare}$), where, however, 
$\alpha_{\rm HLS} =g^2(m_\rho)/4\pi$ is rather large 
$\sim 1$ in the
real-life QCD, $\Delta m_\pi^2 /\Lambda^2 \sim 0.001$, in contrast to
the setting of the little Higgs for the natural hierarchy, 
$\Delta m_\pi^2 /\Lambda^2 \sim (100 {\rm GeV}/ {\rm 10 TeV})^2
 \sim 0.0001$, which
corresponds to $\alpha_{\rm HLS} \sim 0.1$.

At first sight it looks rather difficult to have weakly coupled gauge
theory 
of composite $\rho$ induced by the underlying strong coupling gauge
theory. 
However, it was recently found~\cite{HY:VM,HY:PRep}
 that the $\rho$ gauge coupling becomes
vanishingly small, $\alpha_{\rm HLS} \rightarrow 0$, when we increase 
$N_f$ ($< 11 N_c/2$) from $3$ to  a certain critical point 
$N_f^{\rm crit}$ 
where the chiral symmetry in the underlying QCD
was shown to get restored
in various approaches including the 
lattice simulation~\cite{IKKSY:98}, Schwinger-Dyson 
equation~\cite{Appelquist-Terning-Wijewardhana},
etc. (``Large $N_f$ QCD''). 
Accordingly the $\rho$ mass goes to zero at the critical point
and hence the (longitudinal) $\rho$
becomes the chiral partner of the NG boson 
$\pi$, which we called ``Vector Manifestation (VM)'' of the Wigner
realization of chiral symmetry~\cite{HY:VM,HY:PRep},
characterized by
\begin{eqnarray}
&& F_\pi^2(0) \rightarrow 0 \ , \quad
m_\rho^2 \rightarrow m_\pi^2=0\ , \quad
a(0)
\rightarrow 1 \ .
\label{VM def}
\end{eqnarray}
Through the Wilsonian matching, the chiral restoration in the underlying large
$N_f$ QCD actually dictates that 
the bare parameters of the HLS model should take the following
conditions called ``VM conditions''~\cite{HY:VM,HY:PRep}:
\begin{eqnarray}
 g(\Lambda) \rightarrow 0 \ , 
\quad
a(\Lambda) 
\rightarrow 1 
\label{vector condition:a}
\ ,
\end{eqnarray}
which coincide with the Georgi's vector limit, plus
$F_\pi^2(\Lambda) \rightarrow (F_\pi^{\rm crit})^2 
=\frac{N_f^{\rm crit}}{2(4\pi)^2}
\Lambda^2$,
with $N_f^{\rm crit}\simeq 5.0 \frac{N_c}{3}$ being
estimated
through OPE in the underlying QCD.~\cite{footnote4}

Since $(a,g)=(1,0)$ is a fixed point of the RGEs,
we have $\omega(0)=\omega(m_\rho\rightarrow 0)=\omega(\Lambda)$,
where $\omega(\Lambda)$ is given by 
Eq.~(\ref{match om})\cite{footnote5}:
$\omega(\Lambda) \sim \langle \bar q q \rangle^2/F_\pi^2(0)$ which is
expected to vanish near the critical point, 
since $\langle \bar q q \rangle^2 \sim m^{6-2\gamma_m}$
and $F_\pi(0) \sim m$ near the critical point, where
$\gamma_m$ is the anomalous dimension and $m (\rightarrow 0)$ 
the dynamical mass of the fermion in the underlying large $N_f$ QCD. 
Actually, we 
expect~\cite{Appelquist-Terning-Wijewardhana}
that the large $N_f$ QCD becomes a walking
gauge theory~\cite{walking} near the critical point,
which implies $\gamma_m \simeq 1$. Thus we have
\begin{equation}
\Delta m_\pi^2 / F_\pi^2(0) \sim \alpha_{\rm em} 
\langle \bar q q \rangle^2/F_\pi^4(0)
\sim m^{2-2\gamma_m} \rightarrow c\, ,
\end{equation}
where $c=0 \, (\gamma_m<1)$, and $c \simeq 0.024\ll 1 
\,(\gamma_m=1)$ if estimated through a 
simple ansatz about the $N_f$ dependence 
made in Ref.~\cite{HY:PRep}. Thus the desired hierarchy in the little Higgs
can naturally be realized near the critical point of strongly coupled underlying 
gauge theory.

We would like to thank Andy Cohen, Howard Georgi and Michio Hashimoto
for useful discussions. 
The work is supported in part by the JSPS Grant-in-Aid for the
Scientific Research (B)(2) 14340072.


\begin{thebibliography}{99}
\bibitem{footnote1}
  In the usual calculation of 
  $\Delta m_\pi^2$~\cite{Das},
  the gauge for the photon is 
  taken as 
  $g^{\mu\nu}-4 q^\mu q^\nu /q^2$
  in order to drop the ``seagull term'' proportional to
  $g^{\mu\nu}$. However, 
  in the HLS model we computed $\Delta m_\pi^2$ in arbitrary
  (covariant) gauge,
  the result being gauge-independent as it should.
\bibitem{footnote2}
  This is compared with the conventional formula~\cite{Das}, 
  $(3 \alpha_{\rm em}/4\pi) m_\rho^2 \cdot 2\,\ln (m_{a_1}^2/m_\rho^2)
  \simeq 1432 \, {\rm MeV}^2$, 
  where we put $m_{a_1}^2 \simeq 2 m_\rho^2 (\simeq \Lambda^2)$.
\bibitem{footnote3}
 We have checked 
  the values of these bare parameters for several choices of
  $\Lambda$ and $\Lambda_{\rm QCD}$,
  the result being fairly stable against these changes.
\bibitem{footnote4}
  The RGE for $F_\pi^2$ is readily solved under 
  Eq.~(\ref{vector condition:a})~\cite{HY:duality,HY:VM}
  as 
  $F_\pi^2(0) = (F_\pi^{\rm crit})^2 -\frac{N_f}{2(4\pi)^2}\Lambda^2
   \simeq \frac{N_f}{2(4\pi)^2}\Lambda^2
    \left(N_f^{\rm crit}/N_f - 1\right)$, 
  where 
  the second term is the quadratic divergence whose  
  ``fine tuning'' to
  get a hierarchy more than the naive dimensional
  analysis, $F_\pi^2(0)/\Lambda^2 \ll (1/4\pi)^2$,
  may be naturally realized, if we arrange the $N_f$ in 
  the underlying theory as $N_f^{\rm crit}/N_f - 1\ll 1$.
\bibitem{footnote5}
  The quantum corrections of the form in Eq.~(\ref{PNGmass}) 
  vanish much faster than
  the bare term $\omega(\Lambda)$, since $\alpha_{\rm HLS}$ vanishes as 
  $\langle \bar q q\rangle^2$~\cite{HY:PRep}.
\end{thebibliography}
\end{document}